
\FRONTPAGE
\line{\hfill BROWN-HET-935}
\line{\hfill DAMTP 94-6}
\line{\hfill February 1994}
\vskip1.0truein
\titlestyle{{{\bf COSMIC STRINGS AND ELECTROWEAK BARYOGENESIS}}
\foot{Work supported in part by the Department of Energy under
contract DE-FG02-91ER40688-Task A}}
\bigskip
\author {Robert Brandenberger,$^{1)}$ Anne-Christine Davis,$^{2)}$ Mark
Trodden$^{1)}$ }
\vskip .25in
\item {1)} {\it Physics Department, Brown University, Providence, RI
02912, USA}
\item {2)} {\it Department of Applied Mathematics and Theoretical
Physics and Kings College, University of Cambridge, Cambridge CB3 9EW,
U.K.}
\bigskip
\abstract

The electroweak symmetry is unbroken in the core of cosmic strings originating
from a symmetry breaking at an energy higher than the electroweak scale
$\eta_{EW}$.  The dynamics of such strings may generate a baryon asymmetry
below the electroweak symmetry breaking scale.  This mechanism for electroweak
baryogenesis is most efficient if the scale of string formation is only
slightly higher than $\eta_{EW}$ \eg \ in theories of dynamical symmetry
breaking) and if the strings are superconducting.  The mechanism is also
effective if the electroweak phase transition is second order.

\endpage
{\bf \chapter{Introduction}}

Electroweak barogenesis (for recent reviews see \eg \ Refs. 1 \&2) has become a
topic of much recent activity.  In this paper we suggest a new scenario for
electroweak baryogenesis which also works in models in which the electroweak
phase transition is second order.  Our mechanism makes use of topological
defects (in particular cosmic strings) left behind after a previous phase
transition.\footnote{F1)}{The role of such cosmic strings during the
electroweak phase transition has recently been studied in Ref. 3}.

Until a few years ago, the most popular scenario for baryogenesis$^{4)}$
involved producing the net baryon asymmetry via the out of equilibrium decay of
superheavy gauge and scalar particles after a grand unified symmetry breaking.
However, it was subsequently$^{5)}$ realized that nonperturbative transitions
in the standard electroweak theory will erase any primordial baryon asymmetry
at about the electroweak scale, $\eta_{EW}$, unless $B$ - $L$ is nonvanishing,
$B$ and $L$ denoting baryon and lepton number respectively.  Hence the
nonvanishing baryon to entropy ratio must be due to processes occurring after
$\eta_{EW}$.

As first pointed out by Sakharov$^{6)}$, three conditions need to be satisfied
in order to generate a net baryon number.  First, baryon number violating
processes must exist; second, these processes must violate $C$ and $CP$; and
third, they must occur out of thermal equilibrium.

In the standard electroweak theory all three conditions are satisfied.
Sphalerons transitions$^{7)}$ violate baryon number.\footnote{F2)}{Sphalerons
are gauge and Higgs field configurations which interpolate between equivalent
vacua of the electroweak theory with different baryon number$^{8)}$ (See Ref. 9
for an introductory review).}  The discrete symmetry $CP$ is violated by the
Kobayashi-Maskawa$^{10)}$ mass matrix (however, in all present models of
electroweak baryogenesis the violating effects are too small and must be
enhanced by adding extra $CP$ violation to the Higgs sector).  Finally, out of
equilibrium field configurations may result as remnants of the phase
transition.

The key issue is how the ``out of equilibrium" condition is realized.  In most
previous work$^{1,2,11-13)}$, use was made of bubble walls which form if the
electroweak phase transition is first order.  The bubbles of true vacuum
(broken symmetry) expand in a surrounding sea of false vacuum and then collide.
 The bubble walls are where the net baryon asymmetry is generated since the
baryon number violating processes are unsuppressed, the extra $CP$ violation in
the extended Higgs sector has maximal strength, and since the bubble dynamics
ensures that the processes are out of equilibrium.

However, at present it is unclear$^{14)}$ whether the electroweak phase
transition is sufficiently strongly first order for baryogenesis mechanisms
involving bubble walls to be effective.  In Ref.15 it was pointed out that
topological or nontopological defects may play a role similar to bubble walls
in triggering electroweak baryogenesis (topological defects can also play a
role in GUT--scale baryogenesis$^{16)}$).

During electroweak symmetry breaking no topological defects are produced.
Hence, in Ref.15 we considered the role of electroweak strings$^{17)}$,
nontopological string configurations which can be embedded in the
Weinberg-Salam theory.  However$^{18)}$, such strings are only metastable for
unphysical values of the Weinberg angle $\theta_W$.  In this paper we discuss a
scenario for electroweak baryogenessis which is more robust.  It works for all
valaues of $\theta_W$, and independently of whether the phase transition is
first or second order.  Our mechanism makes use of topological defects produced
in a previous phase transition.

\medskip
{\bf \chapter{Electroweak Baryogenesis with a Second Order Phase Transition}}

Let us review how Sakharov's conditions are realized in electroweak
baryogenesis scenarios and compare the implementations in first and second
order phase transitions.

As mentioned in the introduction, baryon number violation occurs via sphaleron
transitions.  The transition rate is exponentially suppressed in the broken
phase.  However, in the symmetric phase transitions are copious.  Their rate
per unit volume is$^{19)}$
$$
\Gamma_B \sim \alpha_W^4 T^4\eqno\eq
$$
where $\alpha_W = g^2/4\pi$, $g$ being the gauge coupling constant.

We consider extensions of the standard electroweak theory with nonminimal Higgs
structure containing explicit $CP$ violation.  We assume -- as in previous
implementations of electroweak baryogenesis -- that there is a $CP$ violating
phase whose value changes by an amount $\Delta\theta$ during the phase
transition.

In Fig.1 we compare the ways in which the out of equilibrium condition is
realized in models with first and second order phase transitions.  The key role
is played by expanding bubble walls and contracting topological defects
respectively.  In the scenarios of Refs.1,2,11-13, baryogenesis takes place in
the outer edge\footnote{F3)}{According to the numerical simulations of Ref. 20,
out of equilibrium baryon number violating processes will also occur in the
rest of the bubble wall.} of the bubble wall, \ie \ where
$$
\vert\phi \vert < g\,\eta_{EW}\eqno\eq
$$
$\vert\phi\vert$ being the order parameter of the transition.  The amplitude
$\vert\phi\vert$ is increasing at any point in space which the bubble wall
crosses, and hence $CP$ violation has a preferred sign.  Finally, as long as
the bubble wall moves at relativistic speed, there will be no time to establish
thermal equilibrium inside the walls.

With a second order phase transition, the role of the bubble wall is played by
the topological defects.  Baryogenesis takes place inside the core of the
defect where (2.2) is satisfied.  If the defects contract (and eventually
evaporate), then there will be an overall increase in $\vert\phi\vert$, and
hence net baryon number generation.  The field configurations within
contracting topological defects are out of thermal equilibrium.

The standard electroweak theory does not admit any topological defects since
the homotopy groups of the vacuum manifold {\cal M} associated with the
symmetry breaking
$$
SU(2) \times U(1) \longrightarrow U(1)\eqno\eq
$$
are trivial
$$
\Pi_i ({\cal M}) = {\bf 1} \qquad i = 0,1,2 \, .\eqno\eq
$$

In order to obtain topological defects -- cosmic strings to be specific -- we
assume that at an energy scale $\eta$ larger than $\eta_{EW}$ there is another
symmetry breaking which produces strings.  One possibility is to embed $SU(2)
\times U(1)$ in some larger simply connected group $G$ such that at a scale
$\eta$
$$
G\longrightarrow SU(2) \times U(1)\eqno\eq
$$
and
$$
\Pi_1 \left( G/ (SU(2) \times U(1) \right) \not= {\bf 1}\eqno\eq
$$
A second possibility is to assume that electroweak symmetry breaking is induced
dynamically by having a technifermion condensate form at the scale $\eta_{EW}$:
$$
\langle \bar{\psi}_{TC} \psi_{TC} \rangle \not= 0 \, , \,\,\,\, T\leq \eta_{EW}
\, .
\eqno\eq
$$
Here, $\psi_{TC}$ denotes the technifermion.  In this case we can assume that
fermion masses are induced by a second phase transition in the technifermion
sector at a scale $\eta$ which in general is only slightly higher than
$\eta_{EW}$ (for reviews of dynamical symmetry breaking see \eg \ Ref.21).  It
is possible that strings form in this transition.  In the core of these strings
the fermion condensates vanish, the electroweak symmetry is unbroken, and hence
baryon number violating processes are unsuppressed.
\medskip
{\bf \chapter{Strength of the Mechanism}}

In this section we will give a rough estimate of the baryon to entropy ratio
which can be generated using the proposed mechanism, and we will compare the
result with that which may be obtained by the mechanisms of Refs.11-13 which
rely on bubble wall expansion. To simplify the calculations we assume that the
phase transition is rapid, that the strings move relativistically (in order
that the out-of-equilibrium condition is satisfied), and that baryon number
violating processes in the broken phase are suppressed immediately below
$\eta_{EW}$ (in order that the baryon number produced by strings is not washed
out).

The important parameters in our calculation are the total volume $V$, the
volume $V_{BG}$ in which net baryon number violating processes are taking
place, the rate $\Gamma_B$ of these processes (see (2.1)), and the net change
$$
\Delta \theta = \int dt \dot{\theta}\eqno\eq
$$
in the $CP$ violating phase $\theta$. We are making the plausible assumption
that the electroweak symmetry is restored inside the core of the string. In
this case, the mean value of the CP violating phase vanishes in the core. In
the broken phase, the distinguished value of $\theta$ will be nonvanishing.
Hence, for points in space initially inside the string core, the net change in
$\theta$ will have a preferred direction. In this respect there is no
difference between our mechanism and the ones of Refs. 11 - 13.

The net baryon number density $\Delta n_B$ is then given by$^{22)}$
$$
\Delta n_B = {1\over V} \, {\Gamma_B\over T} \, V_{BG} \Delta \theta \,
.\eqno\eq
$$
The volume $V_{BG}$ is determined by the mean separation $\xi$ of the strings
and the radius $R_s$ of the string core (strictly speaking the part of the core
where (2.2) is satisfied and hence $n_B$ violating precesses are unsuppressed).
According to Ref. 11 and some recent work$^{23)}$ there is an extra factor
proportional to ${m_t} / T$ (where $m_t$ is the mass of the top quark) which
decreases the net rate of baryon number violation in the bubble wall (for first
order phase transition mechanisms) and in the cosmic string core (in our
mechanism). Since this factor enters all calculations in the same way, it does
not effect the ratio of strengths of the various mechanisms, and hence we shall
not further consider it in this paper.

The key to the calculation is a good estimate of $V_{BG}$.  Note that the
translational motion of a topological defect does not lead to any net
baryogenesis since $\Delta\theta = 0$ integrated over time.  At the leading
edge of the moving defect, a baryon number with one sign will be produced, but
at the trailing edge baryogenesis will have the opposite sign.  Contraction, on
the other hand, does produce a net $\Delta n_B$.  There is a net
$\Delta\theta\not= 0$ in the entire volume corresponding to the initial string
configuration.  We focus on string loops.  Their mean separation is $\xi(t)$.
Hence, in one horizon volume
$$
V = {4\pi\over 3} \, t^3 \, \eqno\eq
$$
the corresponding volume where net baryon number generation takes place is
$$
V_{BG} \sim R_s^2 \xi (t) \left( {t\over \xi(t)} \right)^3 \, .\eqno\eq
$$
The last factor on the right hand side is the number of string loops per
horizon volume, the second factor is the length of a loop.

Most of the contribution to the baryon to entropy ratio is generated
immediately after $\eta_{EW}$.  Thus, to obtain an order of magnitude estimate
of the strength of our baryogenesis mechanism we will evaluate all quantities
at $t_{EW}$.  Combining (3.2)-(3.4) and (2.1) yields
$$
\Delta n_B (t_{EW} ) \sim \alpha_W^4 \Delta \theta \, \left( {R_s\over
\xi(t_{EW}) }\right)^2 \, T_{EW}^3 \eqno\eq
$$
or
$$
{\Delta n_B\over s} \sim {g^*}^{-1} \alpha_W^4 \Delta \theta \left( {R_s\over
\xi (t_{EW})} \right)^2 \equiv {g^*}^{-1} \alpha_W^4 \Delta \theta \, (SF) \,
,\eqno\eq
$$
with $g^*$ being the number of spin degrees of freedom in radiation, and with
$$
(SF) = \left( {R_s\over \xi (t_{EW})} \right)^2 \, .\eqno\eq
$$
Apart from the factor $(SF)$, this is the same order of magnitude as obtained
in the mechanisms using a first order phase transition$^{11-13)}$.  Hence, we
call $(SF)$ the ``suppression factor".

Above, we implicitly assumed that all strings have the same radius.  This is a
good approximation for strings in the friction dominated epoch$^{24)}$, but not
for a string network in the scaling regime$^{25)}$.  In the latter case (see
Section 4) we need to integrate over all loop sizes to obtain $(SF)$.

The above analysis does not depend on the topology of the defect in a key way.
For collapsing domain walls, our mechanism is stronger since the suppression
factor $(SF)$ would be
$$
(SF) \sim \, {R_c\over \xi (t_{EW})} \, ,\eqno\eq
$$
$R_c$ being proportional to the domain wall core radius.  For collapsing
monopoles, however, the mechanism is weaker since
$$
(SF) \sim \left( {R_c\over \xi(t_{EW} )} \right)^3\, .\eqno\eq
$$

There are two ways to increase $(SF)$:  either we decrease $\xi (t_{EW} )$ or
we increase $R_s$.  The obvious way to decrease $\xi(t_{EW})$ is to decrease
the scale $\eta$ of the string producing phase transition.  The earlier we are
in the friction dominated epoch, the closer the strings are relative to the
horizon since$^{24)}$
$$
\xi (t) \sim \xi (t_f ) \, \left( {t\over t_f }\right)^{5/4} \, , \eqno\eq
$$
$t_f$ being the time of string formation (given by $\eta$).  According to the
Kibble mechanism$^{26)}$
$$
\xi (t_f ) \simeq \lambda^{-1}\,\eta^{-1} \, ,\eqno\eq
$$
where $\lambda $ is the string scalar field   self coupling constant.  The
radius $R_s$ is
$$
R_s \simeq \lambda^{-1/2} g\, \eta^{-1} \, ,\eqno\eq
$$
and therefore
$$
(SF) \sim \lambda \, g^2 \left( {\eta_{EW}\over \eta} \right)^5 \, .\eqno\eq
$$

The second way to increase $(SF)$ is to make the strings superconducting.  As
discovered in Refs.27 \& 28, a supercurrent on the string will lead to a larger
radius $R_s$ of electroweak symmetry restoration.  The maximal current on a
bosonic superconducting cosmic string formed at scale $\eta$ is$^{29)}$
$$
I_{\max} = \eta \, .\eqno\eq
$$
As shown in Refs.25 \& 26, for maximal current the radius $R_s$ becomes
$$
R_s \sim {\eta\over \eta_{EW}^2} \, , \eqno\eq
$$
and hence the maximal value of $(SF)$ is
$$
(SF) \sim \lambda^2 \left( {\eta_{EW}\over \eta} \right) \, . \eqno\eq
$$

In general, however, the initial current on a superconducting cosmic string
will be much less than $\eta$, and therefore $R_s$ will be smaller than the
value given in (3.15).  An optimistic estimate for the current $I$ can be
obtained by assuming that at the time of formation, the winding number $N_s$ of
a loop of size $\xi (t_f )$ is of the order 1, and that the winding of loops at
a later time $t$ (which are produced by the merging of smaller loops) is given
by
$$
N_s (t) = {\cal N}^{1/2} \, N_s (t_f ) \sim {\cal N}^{1/2} \, ,\eqno\eq
$$
where ${\cal N}$ is the ratio of comoving volumes corresponding to the loop
sizes at $t$ and $t_f$ respectively.

A winding number $N_s$ (about the string loop) in the scalar field giving rise
to superconductivity induces a current
$$
I \sim {N_s\over R} \eqno\eq
$$
on a loop of radius $R^{29)}$.  From the analysis of Refs.27 \& 28 it follows
that
$$
R_s \sim I \, .\eqno\eq
$$
Applying (3.17)--(3.19) to loops of radius $\xi (t)$ we obtain
$$
{\cal N}^{1/2} (t) = \left( {\xi (t)\over \xi(t_f )} \, \left( {t_f\over t}
\right)^{1/2} \right)^{3/2} = \left( {\eta\over T}\right)^{9/4} \, ,\eqno\eq
$$
$$
I(t) \sim \lambda \eta \left( {T\over \eta} \right)^{1/4} \sim I_{\max} \left(
{T\over\eta} \right)^{1/4} \eqno\eq
$$
and
$$
R_s \sim \left( {T\over\eta}\right)^{1/4} \left( {\eta\over\eta_{EW}^2} \right)
\, .\eqno\eq
$$
Inserting into (3.7) and using (3.16) we hence get the following estimate for
$(SF)$.
$$
(SF) \sim \lambda^2 \left( {\eta_{EW}\over \eta} \right)^{3/2} \eqno\eq
$$
For a self compling constant $\lambda \sim 1$ and for $\eta$ close to
$\eta_{EW}$, the strength of our mechanism of baryogenesis is hence comparable
to what can be obtained in a first order transition.

\medskip
{\bf \chapter{Two Models}}

We will now briefly discuss two ideas for implementing the above mechanism of
baryogenesis.  The first idea is motivated by dynamical symmetry breaking, the
second is a GUT model.

As is clear from (3.13), if $\eta$ is comparable to $\eta_{EW}$, our
baryogenesis scenario is effective even without strings being superconducting.
Models with a second symmetry breaking at a scale $\eta$ slightly larger than
$\eta_{EW}$ occur automatically in dynamical symmetry breaking
scenarios$^{21)}$.  In such models electroweak symmetry breaking at scale
$\eta_{EW}$ is triggered by a technifermion condensate forming.  In order to
give the ordinary fermions a mass, a previous symmetry breaking is required.
This is sometimes achieved by a larger symmetry called extended technicolor
breaking down to technicolor symmetry at a scale $\eta > \eta_{EW}$.

We assume that, whatever the specifics of this transition, it produces cosmic
strings.  Unless $\eta /\eta_{EW} \gg 1$, the string network will still be
friction dominated at $\eta_{EW}$, and hence the previous calculations apply.
The suppression factor $(SF)$ is given by (3.13) or (3.23) depending on whether
the strings are superconducting or not.,

The second scenario in which our mechanism can be realized is a GUT model which
produces superconducting cosmic strings at a scale $\eta \sim 10^{16}$ GeV.
Note, however, that such scenarios are very highly constrained by observations.
 There is a chance that such strings might  stabilize (become ``springs") at a
current $I_{{\rm spring}} < I_{\max}$$^{30)}$.  In this case the stable
remnants would overclose the Universe$^{31)}$.
 If $I_{{\rm spring}} > I_{\max}$, then electromagnetic radiation from the
strings would lead to spectral distortions in the cosmic microwave background
in excess of the limits set by COBE$^{32)}$.

Setting aside these concerns, we will briefly study the modifications to the
previous calculations required to compute $(SF)$ is this scenario.  The main
difference is that string loops at $T = \eta_{EW}$ are now in the scaling
regime$^{25)}$.  There will be a distribution of loops with radii $R< t_{EW}$
with number density
$$
n(R,t) = \cases {\nu R^{-5/2} t^{-3/2} & \quad$\gamma t < R < t$\cr
\nu \gamma^{-5/2} t^{-4} & \quad $R < \gamma t$\cr}\eqno\eq
$$
where $\gamma \ll 1 $ is a constant determined by the strength of
electromagnetic radiation from the string.  Loops with radius $R=\gamma t$
decay in one Hubble expansion time.  In the above, we are assuming that
electromagnetic radiation dominates over gravitational radiation.  If this is
not the case, then $\gamma$ must be replaced by $\gamma_g \, G \mu $, $\mu$
being the mass per unit length of the string $(\mu \simeq\eta^2)$ and $\gamma_g
\sim 100$ $^{33)}$. In other words, $\gamma$ is bounded from below
$$
\gamma > \gamma_g G \mu\, .\eqno\eq
$$

The volume $V_{BG}$ in which net baryogenesis takes place can be estimated by
integrating over all string loops present at $t_{EW}$
$$
{V_{BG}\over V} \simeq 2 \pi \int_0^{\gamma t} \, d R \, R\, R_s^2 \, n \left(
R,t_{EW} \right) = \pi \, \nu\, \gamma^{-1/2} \left( {R_s\over t_{EW}
}\right)^2 \eqno\eq
$$
and hence
$$
{\Delta n_B\over s} \sim \alpha_W ^4 \Delta \theta \nu \gamma^{-1/2} \left(
{R_s\over t_{EW} } \right)^2 \equiv \alpha_W ^4 \Delta \theta \, (SF) \eqno\eq
$$
with
$$
(SF) = \nu\gamma^{-1/2} \left( {R_s\over t_{EW}} \right)^2 \eqno\eq
$$
Without superconductivity, the rate of baryogenesis is completely negligible.
With (3.12), the result (4.5) becomes
$$
(SF) \sim  \nu\gamma^{-1/2} g^2 \lambda^{-1} \left( {T_{EW}^2\over \eta m_{pl}
} \right)^2 \sim 10^{-62} \, , \eqno\eq
$$
$m_{pl}$ being the Planck mass.  For the numerical estimate we have used $\nu
\sim 10^{-2}$, $\lambda \sim g\sim 1$ and $\gamma = \gamma_g G \mu \sim
10^{-4}$.

In GUT scenarios, superconductivity greatly enhances the rate of baryogenesis.
With maximal current $I_{\max}$, it follows by comparing (3.15) and (3.12) that
$(SF)$ is larger by a factor of $(\eta/\eta_{EW} )^4$, \ie
$$
(SF) \sim  \nu\gamma^{-1/2} \left( {\eta\over m_{pl}} \right)^2 \sim 10^{-8}
\eqno\eq
$$
with $\nu \sim 10^{-2}$ and $\gamma \sim 1$.  Even this result is much smaller
than the value
$$
(SF) > 10^{-2} \eqno\eq
$$
required in order that the mechanism can explain the observed baryon to entropy
ratio.  In fact, even the estimate (4.7) is too optimistic since strings are
unlikely to have maximal current.

We conclude that cosmic string mediated baryogenesis is much more efficient if
the strings are produced only shortly before $\eta_{EW}$.

\medskip
{\bf \chapter{Conclusions}}

We have discussed a new mechanism for electroweak baryogenesis which operates
even if the electroweak phase transition is second order.  Topological defects
produced in a previous phase transition can play a role analogous to bubble
walls in generating a net baryon asymmetry.  More specifically, we have
analyzed models with cosmic strings left over from a phase transition at a
scale $\eta >  \eta_{EW}$.  Provided that $\eta/\eta_{EW}$ is of the order 1
(as is the case in some technicolor models), it is possible to generate a net
baryon to entropy ratio of the order $n_B/s \sim 10^{-10}$, the required value
to match observations.

\medskip

\noindent{\bf Acknowledgements}

Two of us (R.B \& M.T.) wish to thank the particle theory group at DAMTP for
hospitality. We are grateful to Neil Turok for valuable comments on the first
draft of this paper. Our work was supported in part by the US Department of
Energy under Grant DE-FG02-91ER40688, Task A and by an NSF-SERC Collaborative
Research Award NSF-INT-9022895 and SERC GR/G37149.
\bigskip

\noindent{\bf Figure Caption}
\item{} A comparison between the electroweak baryogenesis mechanisms using
first order phase transitions (top) and our mechanism (bottom). The contracting
topological defect (bottom) plays a role similar to that of the expanding
bubble wall (top) in that it is the location of extra CP violation and of
baryon number violating processes taking place out of equilibrium.
\bigskip

\noindent{\bf References}

\medskip
\pointbegin
N. Turok, in `Perspectives on Higgs Physics', ed. G. Kane (World Scientific,
Singapore, 1992).
\point
A. Cohen, D. Kaplan and A. Nelson, {\it Ann. Rev. Nucl. Past. Sci.} {\bf 43},
27 (1993).
\point
S. Duari and U. Yajnik, `Cosmic Strings at the Electroweak Phase Transition',
{\it Phys. Lett. B}, in press (1994).
\point
S. Dimopoulos and L. Susskind, {\it Phys. Rev.} {\bf D18}, 4500 (1978); M.
Yoshimura, {\it Phys. Rev. Lett.} {\bf 41}, 281 (1978); A. Ignatiev, N.
Krasnikov, V. Kuzmin and A. Tavkhelidze, {\it Phys. Lett.} {\bf B76}, 436
(1978); S. Weinberg, {\it Phys. Rev. Lett.} {\bf 42}, 850 (1977); D. Toussaint,
S. Trieman, F. Wilczek and A. Zee, {\it Phys. Rev.} {\bf D19}, 1036 (1979).
\point
V. Kuzmin, V. Rubakov and M. Shaposhnikov, {\it Phys. Lett.} {\bf B155}, 36
(1985); P. Arnold and L. McLerran, {\it Phys. Rev.} {\bf D36}, 581 (1987).
\point
A. Sakharov, {\it Pisma Zh. Ekop. Teor. Fiz.} {\bf 5}, 32 (1967).
\point
N. Manton, {\it Phys. Rev.} {\bf D28}, 2019 (1983); F. Klinkhamer and N.
Manton, {\it Phys. Rev.} {\bf D30}, 2212 (1984).
\point
G. t'Hooft, {\it Phys. Rev. Lett.} {\bf 37}, 8 (1976).
\point
P. Arnold, `Introduction to Baryon Violation in Standard Electroweak Theory',
in TASI'90 (World Scientific, Singapore, 1991).
\point
M. Kobayashi and M. Maskawa, {\it Prog. Theor. Phys.} {\bf 49}, 652 (1973).
\point
N. Turok and T. Zadrozny, {\it Phys. Rev. Lett.} {\bf 65}, 2331 (1990); N.
Turok and J. Zadrozny, {\it Nucl. Phys.} {\bf B358}, 471 (1991); L. McLerran,
M. Shaposhnikov, N. Turok and M. Voloshin, {\it Phys. Lett.} {\bf B256}, 451
(1991).
\point
A. Cohen, D. Kaplan and A. Nelson, {\it Phys. Lett.} {\bf B263}, 86 (1991).
\point
A. Nelson, D. Kaplan and A. Cohen, {\it Nucl. Phys.} {\bf B373}, 453 (1992).
\point
M. Dine, R. Leigh, P. Huet, A. Linde and D. Linde, {\it Phys. Rev.} {\bf D46},
550 (1992).
\point
R. Brandenberger and A. Davis, {\it Phys. Lett.} {\bf B308}, 79 (1993).
\point
R. Brandenberger, A. Davis and M. Hindmarsh, {\it Phys. Lett.} {\bf B263}, 239
(1991); A.-C. Davis and M. Earnshaw, {\it Nucl. Phys.} {\bf B394}, 21 (1993).
\point
T. Vachaspati, {\it Phys. Rev. Lett.} {\bf 68}, 1977 (1992).
\point
M. James, L. Perivolaropoulos and T. Vachaspati, {\it Nucl. Phys.} {\bf B395},
534 (1993).
\point
T. Ambjorn, M. Laursen and M. Shaposhnikov, {\it Phys. Lett.} {\bf B197}, 49
(1989); T. Ambjorn, T. Askgaard, H. Porter and M. Shaposhnikov, {\it Nucl.
Phys.} {\bf B353}, 346 (1991).
\point
D. Grigoriev, M. Shaposhnikov and N. Turok, {\it Phys. Lett.} {\bf B275}, 395
(1992).
\point
See \eg \ E. Farhi and L. Susskind, {\it Phys. Rep.} {\bf 74}, 277 (1981).
\point
M. Dine, O. Lechtenfeld, B. Sakita, W. Fischler and J. Polchinski, {\it Nucl.
Phys.} {\bf B342}, 381 (1990).
\point
G. Giudice and M. Shaposhnikov, `Strong Sphalerons and Electroweak
Baryogenesis', CERN preprint CERN-TH 7080/93 (1993); M. Dine and S. Thomas,
`Electroweak Baryogenesis in the Adiabatic Limit', Santa Cruz preprint SCIPP
94/01 (1994); M. Joyce, T. Prokopec and N. Turok, `Why Hypercharge Doesn't Make
Baryons', Princeton preprint PUP-TH-1436 (1993).
\point
T. W. B. Kibble, {\it Acta Phys. Pol.} {\bf B13}, 723 (1982); A. Everett, {\it
Phys. Rev.} {\bf D24}, 858 (1981); M. Hindmarsh, Ph. D. Thesis, University of
London, unpublished (1986).
\point
Ya. B. Zel'dovich, {\it Mon. Nat. R. Astr. Soc.} {\bf 192}, 663 (1980); A.
Vilenkin, {\it Phys. Rev. Lett.} {\bf 46}, 1169 (1981); N. Turok and R.
Brandenberger, {\it Phys. Rev.} {\bf D37}, 2075 (1986).
\point
T. W. B. Kibble, {\it J. Phys.} {\bf A9}, 1387 (1976).
\point
J. Ambjorn, N. Nielsen and P. Olesen, {\it Nucl. Phys.} {\bf B310}, 625 (1988).
\point
W. Perkins and A. Davis, {\it Nucl. Phys.} {\bf B406}, 377 (1993).
\point
E. Witten, {\it Nucl. Phys.} {\bf B249}, 557 (1985).
\point
E. Copeland, M. Hindmarsh and N. Turok, {\it Phys. Rev. Lett.} {\bf 58}, 1910
(1987); R. Brandenberger, A.-C. Davis, A. Matheson and M. Trodden, {\it Phys.
Lett.} {\bf B293}, 287 (1992).
\point
A. Davis and R. Brandenberger, {\it Phys. Lett.} {\bf B284}, 81 (1992).
\point
J. Mather et al., {\it Ap J. (Lett.)} {\bf 354}, L37 (1990).
\point
T. Vachaspati and A. Vilenkin, {\it Phys. Rev.} {\bf D31}, 3052 (1985); N.
Turok, {\it Nucl. Phys.} {\bf B242}, 520 (1984); C. Burden, {\it Phys. Lett.}
{\bf B164}, 277 (1985)
\end